\documentclass[aps,prb,twocolumn,groupedaddress,superscriptaddress,amsmath,amsfonts,amssymb]{revtex4-1}
\usepackage[english]{babel}
\usepackage{ucs}
\usepackage[utf8x]{inputenc}
\usepackage{graphicx}
\usepackage{bm}
\usepackage{bbm}
\usepackage{empheq}

\newcommand{\bra}[1]{\langle #1|}
\newcommand{\ket}[1]{|#1\rangle}

\newcommand{\mm}[1]{\mathrm{#1}}
\newcommand{\ui}{\mathrm{i}}
\newcommand{\ue}{\mathrm{e}}
\newcommand{\ud}{\mathrm{d}}

\newcommand{\abs}[1]{\left|#1\right|}
\newcommand{\drv}[2]{\frac{\ud#1}{\ud#2}}

\begin{document}

\title{Harnessing the GaAs quantum dot nuclear spin bath for quantum control}

\author{Hugo Ribeiro}
\affiliation{Department of Physics, University of Konstanz, D-78457 Konstanz, Germany}
\author{J. R. Petta}
\affiliation{Department of Physics, Princeton University, Princeton, NJ 08544, USA}
\author{Guido Burkard}
\affiliation{Department of Physics, University of Konstanz, D-78457 Konstanz, Germany}



\begin{abstract}
We theoretically demonstrate that nuclear spins can be harnessed to coherently control
two-electron spin states in a double quantum dot. Hyperfine interactions lead to an
avoided crossing between the spin singlet state and the $m_{\mm{s}}=+1$ triplet state,
$\mm{T}_+$. We show that a coherent superposition of singlet and triplet states can be
achieved using finite-time Landau-Zener-Stückelberg interferometry. In this system the
coherent rotation rate is set by the Zeeman energy, resulting in $\sim 1$ nanosecond
single spin rotations. We analyze the coherence of this spin qubit by considering the
coupling to the nuclear spin bath and show that $T_2^{*} \sim 16\,\mm{ns}$, in good
agreement with experimental data. Our analysis further demonstrates that efficient single
qubit and two qubit control can be achieved using Landau-Zener-Stückelberg
interferometry. 
\end{abstract}

\maketitle

\section{Introduction}

Considerable effort has been made in the past few years to implement qubits in nanoscale
solid state structures. One of the most promising candidates are spin qubits confined in
electrostatically defined quantum dots (QDs) embedded in GaAs structures
\cite{loss_divicenzo_proposaL_QD, hanson_review}. A universal set of quantum gates has
been demonstrated in GaAs double quantum dots (DQD) through the achievements of
single-spin rotations and the two-spin exchange interaction that generates the
$\sqrt{\mm{SWAP}}$ gate \cite{koppens_nature2006,petta_science2005,nowack_science2007}.
Despite these advances, coherence times are limited by the hyperfine interaction, which
couples the trapped electron spin in the quantum dot to the spin-$\frac{3}{2}$ nuclei of
the GaAs substrate. The resulting nuclear fields cause rapid electron spin
dephasing, leading to an inhomogenous dephasing time $T_2^*\sim 10\,\mm{ns}$.
As each electron spin is coupled to approximately one million nuclei, the resulting
behavior of the coupled electron-nuclear spin system is complicated and leads to rich
dynamics that are sensitive to experimental parameters \cite{lukin_2010}. 

The hyperfine interaction has traditionally been viewed as a nuisance. However, a recent
experiment demonstrates that generation of a controlled nuclear field gradient can be
used to drive fast spin rotations \cite{foletti_nphys2009}. The development of quantum
control methods in semiconductor quantum dots that are based on nuclear spin interactions
could lead to new paradigms for single spin control.

\begin{figure}
\includegraphics[width=0.48\textwidth]{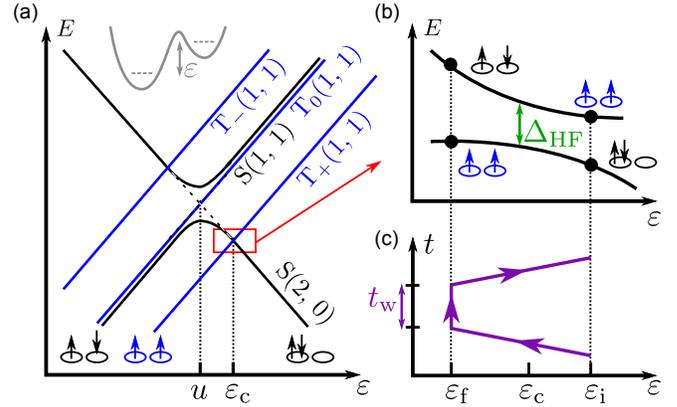}
\caption{(color online) (a) Energy diagram for the relevant states in the DQD as a
function of $\varepsilon$. The spin states for the implementation of the qubit are the
hybridized singlet $\mm{S}$ and the triplet $\mm{T}_+$. (b) A coherent
superposition of the qubit states is generated by LZS interferometry. (c) The initialized
$\mm{S(2,\,0)}$ is swept through the avoided crossing by means of an applied linear gate
voltage pulse $\varepsilon_{\mm{i}} \to \varepsilon_{\mm{f}}$. The final
state is a coherent superposition of $\mm{S}$ and $\mm{T}_+$ generated by LZS tunneling.
At $\varepsilon_{\mm{f}}$, the system evolves in the external magnetic field $B$ for a
time $t_{\mm{w}}$ before a reverse gate voltage pulse brings the system back to
$\varepsilon_{\mm{i}}$ where a QPC measurement is performed to determine 
the singlet state return probability, $P_{\mm{S}}$.}
\label{energy_diagram}
\end{figure}

We theoretically show that hyperfine interactions can be harnessed for quantum control in
GaAs semiconductor quantum dots. In the presence of an external magnetic field $B$, which
splits the triplet states, the hyperfine interaction results in an avoided crossing
between the spin singlet $\mm{S}$ and spin triplet $\mm{T}_{+}$, which form the basis of
a new type of spin qubit. Coherent quantum control of this qubit has already been
experimentally achieved through Landau-Zener-Stückelberg (LZS) interferometry
\cite{petta_science2010}, a technique previously used to coherently manipulate
superconducting qubits \cite{oliver_science2005,Shevchenko_review2009} and which is based
on the interference due to repeated LZS tunneling events \cite{landau32, zener32,
stuckelberg32, majorana32}. The original LZS problem studies a two level system which
exhibits an avoided crossing when an external control parameter is changed. If the
control parameter is time dependent, the system can be brought from an initial state
through the avoided crossing. This passage may result in a change of populations and
relative phase of the states. In the LZS version, the system is driven from $t_{\mm{i}} =
-\infty \to t_{\mm{f}} = \infty$ and the difference in energy between the two states is a
linear function of time, $\Delta E(t) = \alpha t$, which leads to the well-known result
for the non-adiabatic transition probability $P_{\mm{LZS}} = \exp(-2 \pi |\bra{\mm{S}}
H_{\mm{int}}\ket{\mm{T}_+}|^2 /\alpha \hbar)$. 

For the DQD system, the usual infinite-time asymptotic theory describing LZS
interferometry cannot be used. The avoided crossing originates from the hyperfine
interaction between the electronic spins and the nuclear spins whose fluctuations result
in a poorly defined crossing position. The phenomena observed in the experiments can only
be properly described using finite-time LZS theory \cite{vitanov_time_lz}.

To prove that our formalism describes correctly the coherent manipulation of the
$\mm{S}$-$\mm{T}_{+}$ based qubit, we will compare it to the experimentally measurable
quantity $P_{\mm{S}}$, the singlet return probability. We then show how single qubit
operations can be engineered either by the Euler angle method for rotations or by only
using LZS interferometry. Finally, we demonstrate that a two qubit gate can be achieved
by capacitively coupling two $\mm{S}$-$\mm{T}_{+}$ qubits. In contrast to the
$\mm{S}$-$\mm{T}_{0}$ qubit, where the rotation rate is set by a charge-noise-susceptible
exchange energy, the rotation rate in the $\mm{S}$-$\mm{T}_{+}$ qubit is set by the
Zeeman energy and approaches $1\,\mm{ns}$ for modest magnetic fields of $100\,\mm{mT}$. 

\section{Model}

The spin preserving Hamiltonian,
\begin{equation}
H_0 = \sum_{j s} \varepsilon_{js} n_{j s}
+ u\sum_{j} n_{j\uparrow} n_{j\downarrow}
+ \tau\sum_{s}\!\left(c_{1 s}^{\dag} c_{2 s} + \mathrm{h.c}\right)\!,
\label{H_trap}
\end{equation}
with $\varepsilon_{js}= \varepsilon_j + g^{*}\mu_{\mathrm{B}}B s/2$ and $n_{js}=c_{j
s}^{\dag} c_{j s}$ describes the coupling between two electrons in a DQD in a magnetic
field $B$. $g^{*}$ denotes the effective Landé g-factor ($-0.44$ for GaAs),
$\mu_{\mathrm{B}}$ the Bohr magneton, the $j=1,2$ and $s=\uparrow,\downarrow=\pm 1$ label
the dot number and spin. The first term is the single-particle energy of the confined
electrons, the second accounts for the Coulomb energy $u$ of two electrons on the same
QD, and the last for tunneling with strength $\tau$ between the dots. 

The diagonalization of the first two terms of Eq.~\eqref{H_trap} leads to the relevant
charge states of a DQD: the singlets $\mathrm{S}(0,2)$, $\mathrm{S}(2,0)$,
$\mathrm{S}(1,1)$ and triplets $\mathrm{T}_{\pm,0}(1,1)$ where $(l,\,r)$ denotes the
charge configuration of the dots (see Fig.~\ref{energy_diagram}(a)). The other states can
be neglected as they have energies much higher than those considered here. The degeneracy
of the singlets $\mathrm{S}(2,0)$ and $\mathrm{S}(1,1)$ at $\varepsilon = \pm u$ is
lifted by the inter-dot tunneling, resulting in a splitting of $\sqrt{2} \tau$.
The hyperfine interaction $H_{\mathrm{HF}} = \mathbf{S}_1\cdot\mathbf{h}_1 +
\mathbf{S}_2\cdot\mathbf{h}_2$ between the electron spins $\mathbf{S}_i$ and the nuclear
spins $\mathbf{I}_i^k$ opens a splitting $\Delta_{\mm{HF}}$ at the degeneracy point
$\varepsilon_{\mm{c}}$ of the singlet $\mm{S}$ and the triplet $\mm{T}_{+}$. Here,
$\mathbf{h}_{\mathrm{i}} = \sum_{k=1}^{n(i)} A_i^k \mathbf{I}_i^k$ is the Overhauser
(effective nuclear) field operator. The sum runs over the $n(i)$ nuclear spins in dot
$i$ and $A_i^k = v_{ik} \nu_0 \left|\Psi_i (\mathbf{r}_k)\right|^2$ is the hyperfine
coupling constant with the $k$-th nucleus in dot $i$, with $\Psi_i (\mathbf{r}_k)$ the
electron wave function, $\nu_0$ the volume of the unit cell and $v_{ik}$ the hyperfine
coupling strength. From now on, since we assume symmetric dots, we have $n(1) = n(2) =
n$.
Introducing $S_i^\pm=S_i^x\pm \mm{i} S_i^y$ and  $h_i^\pm=h_i^x\pm \mm{i} h_i^y$, we write
\begin{equation}
H_{\mathrm{HF}} = \frac{1}{2}\sum_{i}\left(2 S_i^z h_i^z + S_i^+h_i^- + S_i^-h_i^+\right).
\label{hhf}
\end{equation}

To determine which spin states are relevant for our theory, we consider $P_{\mm{LZS}}$
from the asymptotic LZS model and the result of Ref.~\onlinecite{kayanuma_jpb1985} for multiple
level crossings to estimate the order of magnitude of the transition probabilities. The
initialization of the system is done by preparing a singlet $\mm{S}(2,\,0)$
$(\varepsilon_{\mm{i}} > \varepsilon_{\mm{c}})$, then $\varepsilon$ is swept to achieve
$\varepsilon_{\mm{f}}<\varepsilon_{\mm{c}}$. During this operation, the system goes
through three avoided crossings (cf Fig.~\ref{energy_diagram}). To estimate the matrix
element $|\bra{\mm{S}} H_{\mm{HF}}\ket{\mm{T}_+}|^2 = |\bra{\mm{S}}
H_{\mm{HF}}\ket{\mm{T}_-}|^2 = |\Delta_{\mm{HF}}|^2$, we use the experimentally found
value of $\Delta_{\mm{HF}} = 60\,\mm{neV}$ from Ref.~\onlinecite{petta_science2010}. The matrix element
entering $P_{\mm{LZS}}$ at the avoided crossing between $\mm{S}(2,\,0)$ and the excited
singlet state $\mm{S}^{\prime}(1,\,1)$ is given by $|\bra{\mm{S}(2,\,0)}
H_0\ket{\mm{S}^{\prime}(1,\,1)}|^2 = 2\tau^2$, with $\tau = 5\,\mm{\mu eV}$. The order of
magnitude of $\alpha$ is taken between $10^{-3} - 10^{-2}\,\mm{meV/ns}$. We find
$P_{\mm{T}_+} \simeq 0.97\,-\,0.99$ and  $P_{\mm{S}^{\prime}(1,\,1)},\,P_{\mm{T}_-} \ll
10^{-8}$. These results show that population of the excited singlet and  $\mm{T}_-$ level
are negligible. $\mm{T}_0$ can also be neglected because it does not cross with any other
level, it splits from $\mm{S}(1,\,1)$ due to the exchange coupling \cite{hanson_prl2007}. 

Near the $\mm{S}$-$\mm{T}_+$ crossing, the dynamics can be restricted to the Hilbert
space spanned by $\mm{T}_{+}(1,1)$, $\mm{S}(1,1)$, and $\mm{S}(2,0)$ and described by
\begin{equation}
H_{\mm{S},\mm{T}_+} \simeq
\begin{pmatrix}
g^* \mu_{\mm{B}} B & 0 & 0\\
0 & 0 & \sqrt{2} \tau\\
0 & \sqrt{2} \tau & u - \varepsilon \\
\end{pmatrix}
\label{hst+}
\end{equation}
where we can neglect an additive term $\propto \varepsilon \mathbbm{1}$, with
$\varepsilon = \varepsilon_1 - \varepsilon_2$ the detuning of the dots. According to our
previous estimate, we can reduce Eq.~\eqref{hst+} to a $2\times2$ Hamiltonian which only takes
into account the lowest hybridized singlet $\ket{\mm{S}} = c(\varepsilon)
\ket{\mm{S}(1,\,1)} + \sqrt{1-c(\varepsilon)^2} \ket{\mm{S}(2,\,0)}$ and the triplet
$\ket{\mm{T}_+} = \ket{\mm{T}_+ (1,\,1)}$, where $c (\varepsilon) = \left(-u +
\varepsilon - \eta \right)/ \sqrt{8 \tau^2 + (u - \varepsilon + \eta )^2}$, with $\eta=
\sqrt{8 \tau^2 + (u - \varepsilon)^2}$. The energy associated with the lowest hybridized
singlet is $E_{\mm{S}}(\varepsilon) = \left(u - \varepsilon -\eta \right) /2$ and the
energy of the triplet is $E_{\mm{T}_+}= g^* \mu_{\mm{B}} B$. The Hamiltonian describing
the dynamics of the lowest energy states in the vicinity of $\mm{S}$-$\mm{T}_+$ can
therefore be written as 
\begin{equation}
H_0 (\varepsilon) = E_{\mm{S}} (\varepsilon) \ket{\mm{S}}\bra{\mm{S}} +
E_{\mm{T}_+} \ket{\mm{T}_+}\bra{\mm{T}_+}.
\label{H_0}
\end{equation}

Another relevant quantity derived from Eq.~\eqref{hst+} is the degeneracy position $\varepsilon_{\mm{c}}$
of $\mm{S}$ and $\mm{T}_{+}$, given by the funnel-shaped function
\begin{equation}
\varepsilon_{\mm{c}} (u,\,B) = u + 2 \tau^2 / g \mu_{\mm{B}} B  - g \mu_{\mm{B}} B.
\label{crossing}
\end{equation}

\section{Semi-classical theory}

We model the  Overhauser field classically, such that it acts on the electron spin as a
magnetic field $\mathbf{B}_{\mm{n},\,i}$ with $\mathbf{h}_i = g^* \mu_{\mm{B}}
\mathbf{B}_{\mm{n},\,i}$ and its physical properties are given by a statistical
distribution that reflects the quantum fluctuations of the nuclear ensemble. At typical
operating temperatures and external magnetic fields $k_B T \gg g_{\mm{N}}\mu_{\mm{N}} B$,
where $g_{\mm{N}}$ and $\mu_{\mm{N}}$ are the nuclear $g$ factor and magneton. In this
limit we can assume the nuclei to be completely unpolarized, resulting in a Gaussian
distribution of nuclear fields\cite{khaetskii_e-spin_decoherence,coish_prb2005}
$p(B_{\mm{n},\,i}) = (1/\sqrt{2 \pi} \sigma)\mm{e}^{-B_{\mm{n},\,i}^2 / 2\sigma^2}$, with
$\sigma = A / g^* \mu_{\mm{B}} \sqrt{n}$ and $A \approx 90\,\mm{\mu eV}$. The effective
Hamiltonian describing the qubit dynamics around the $\mm{S}$-$\mm{T}_+$ avoided crossing
is given by
\begin{equation}
H_{\mm{eff}}(\varepsilon)  = H_0(\varepsilon) + \frac{1}{2} g^* \mu_{\mm{B}}\sum_i \left(
S_i^+ B_{\mm{n},\,i}^- + S_i^- B_{\mm{n},\,i}^+ \right) .
\label{h_eff}
\end{equation}
We include the nuclear Zeeman spliting in the $\mm{T}_+$ energy, $E_{\mm{T}_+} = g^*
\mu_{\mm{B}} (B + B_{\mm{n}}^z)$, where $B_{\mm{n}}^z = B_{\mm{n,\,1}}^z +
B_{\mm{n,\,2}}^z$ and $B_{\mm{n},\,i}^{\pm} = B_{\mm{n},\,i}^x \pm \mm{i}
B_{\mm{n},\,i}^y$.
The classical approximation of the Overhauser fields is possible because the nuclear
state changes only slightly after a single sweep \cite{ribeiro_prl2009}.
To obtain an analytical expression of the LZS propagator (see Appendix
\ref{sec:lzsprop}), we linearize the difference in energy $\Delta E(t)$ around $t=0$
\cite{vitanov_time_lz} by assuming $\varepsilon (t) = \gamma t + \varepsilon_{\mm{c}}$,
we find 
\begin{equation}
\alpha = \frac{(g \mu_{\mm{B}} (B + B_{\mm{n}}^z))^2} {2 \tau^2 + (g \mu_{\mm{B}} (B + B^z_{\mm{n}}))^2}\gamma. 
\label{alpha}
\end{equation}

Here $\gamma$ is the rate at which the external voltage gates are ramped. Control over
$\alpha$ can therefore be achieved by modifying $\gamma$. 

To test our model we first compute the singlet return probability $P_{\mm{S}}$, an
experimentally observable quantity, as a function of the final detuning
$\varepsilon_{\mm{f}}$ and waiting time $t_{\mm{w}}$,

\begin{equation}
P_{\mm{S}} = \int \prod_{k=1,2}\mm{d}\mathbf{B}_{\mm{n},k} \,p(\mathbf{B}_{\mm{n},k})
\left| \bra{\mm{S}}U(\mathbf{B}_{\mm{n},\,1},\,\mathbf{B}_{\mm{n},\,2})\ket{\mm{S}}\right|^2,
\label{PS_eff}
\end{equation}
with 
\begin{equation}
U(\mathbf{B}_{\mm{n},1},\mathbf{B}_{\mm{n},2})
=U_{\mm{b}}(\mathbf{B}_{\mm{n},1},\mathbf{B}_{\mm{n},2}) U_{\mm{w}}(B^z_{\mm{n}})
U_{\mm{f}}(\mathbf{B}_{\mm{n},1},\mathbf{B}_{\mm{n},2}),
\label{full_prop}
\end{equation}
where $U_{\mm{b,\,f}} = \mm{T}\exp[-\mm{i} \int_{t_\mm{i}}^{t_{\mm{f}}} \mm{d}t
H_{\mm{eff}}(\varepsilon(t))/\hbar] $ are the backward and forward LZS propagators and
$U_{\mm{w}} \simeq \mm{T}\exp[-\mm{i} \int_{0}^{t_{\mm{w}}} \mm{d}t
H_0(\varepsilon_{\mm{f}})/\hbar]$ describes the evolution of the system during the
waiting time $t_{\mm{w}}$ at the final detuning position $\varepsilon_{\mm{f}}$ with
$\left| \varepsilon_{\mm{f}} - \varepsilon_{\mm{c}}\right| \gtrsim \Delta_{\mm{HF}}$.
\begin{figure}
\includegraphics[width=0.48\textwidth]{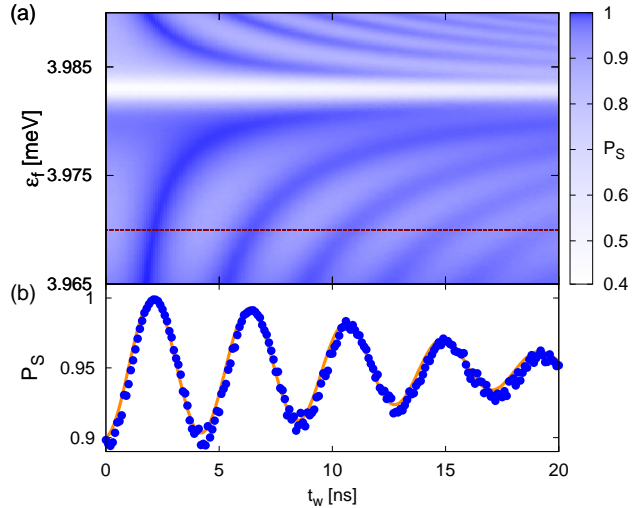}
\caption{(color online) Theoretical results. (a) The singlet return probability
$P_{\mm{S}}$ from the semi-classical model as a function of the waiting time
$t_{\mm{w}}$ and final detuning $\varepsilon_{\mm{f}}$. We
find nanosecond oscillation periods and the dephasing time
$T_2^* \sim 16\,\mm{ns}$, in good agreement with experiment.
We used $B=100\,\mm{mT}$, $\gamma = 0.015\,\mm{meV/ns}$, $u =
4\,\mm{meV}$ and $\tau = 5\,\mm{\mu eV}$. (b) $P_{\mm{S}}$ (blue) as a function of
$t_{\mm{w}}$ for $\varepsilon_{\mm{f}} = 3.97\,\mm{meV}$ (horizontal line in (a)). We
plot (orange) $ C\sin(\omega t) \exp(-(t/T_2^*)^2)$ where $\omega = \left|E_{\mm{S}}
(\varepsilon_{\mm{f}}) - E_{\mm{T}_+}\right| / \hbar \sim 2 \pi \cdot 0.23\,\mm{GHz}$,
$C=0.95$ and $T_2^* = (16.0 \pm 0.4)\,\mm{ns}$ is extracted from a best fit.}
\label{eff_res}
\end{figure}
We evaluate $P_{\mm{S}}$ by numerical sampling. Instead of estimating $\sigma$ from
$n$ and $A$ of the QDs, here we use the experimentally determined
\cite{petta_science2010} $\Delta_{\mm{HF}} = 60\,\mm{neV}$ to derive $\sigma \sim
1.67\,\mm{mT}$. We have $\langle \Delta_{\mm{HF}}^2 \rangle = \langle (\delta
B_{\mm{n}}^-)^2 \rangle = 4 \sigma^2$, with $\delta B^-_{\mm{n}} \equiv g^* \mu_{\mm{B}}( B^-_{\mm{n},\,2} -
B^-_{\mm{n},\,1})/2\sqrt{2} = \bra{\mm{T}_+} H_{\mm{HF}}^\perp \ket{\mm{S}}$. In
Fig.~\ref{eff_res} we show $P_{\mm{S}}$ as a function of $\varepsilon_{\mm{f}}$ and
$t_{\mm{w}}$ for $\gamma = 0.015\,\mm{meV/ns}$, $B = 100\,\mm{mT}$,
$u=4\,\mm{meV}$, and $\tau = 5\,\mm{\mu eV}$. We use a square pulse with a ramping time
$|t_{\mm{f}} - t_{\mm{i}}|$ fixed to $1.5\,\mm{ns}$ and the initial
detuning $\varepsilon_{\mm{i}}$ is varied to reach different values of
$\varepsilon_{\mm{f}}$.

We identify coherent oscillations as a function of $t_{\mm{w}}$ and
$\varepsilon_{\mm{f}}$. From a best fit, we obtain the decoherence time $T_2^* = (16.0
\pm 0.4)\,\mm{ns}$, which agrees well with experiment.  The decoherence is mainly due to
the fluctuations of $B_{\mm{n}}^z$. The period of
the temporal oscillations is $T = h/\left|E_{\mm{S}}(\varepsilon_{\mm{f}}) -
E_{\mm{T}_+}\right|\sim 4.3\,\mm{ns}$ for $\varepsilon_{\mm{f}} = 3.97\,\mm{meV}$ (see
Fig.~\ref{eff_res}(b)). 
For a fixed $B$, a shorter period can be obtained for smaller $\varepsilon_{\mm{f}}$, the
fastest oscillations being defined by the Zeeman energy. To further decrease the period
the external $B$ field can be increased and hence the qubit manipulation could be done in
a time scale of $100\,\mm{ps}$ for $B\sim 1.6\,\mm{T}$, which would allow $\sim 160$
coherent operations within $T_2^*$. In the exchange gate demonstrated in
Ref.~\onlinecite{petta_science2005}, $\mm{d}J/\mm{d}\varepsilon$ increases with
$\varepsilon$, which results in faster dephasing for faster rotations. In contrast, here
the rotation rate is set by the Zeeman energy, which is independent of $\varepsilon$ far
from the $\mm{S}$-$\mm{T}_+$ avoided crossing. As a result, the coherent oscillation
frequency can be increased without making the qubit more susceptible to gate voltage
fluctuations by simply increasing $B$. Far from the avoided crossing the level spacing is
independent of detuning, similar to the ``sweet spot'' in superconducting qubits
\cite{vion_science2002}.

\begin{figure}
\includegraphics[width=0.48\textwidth]{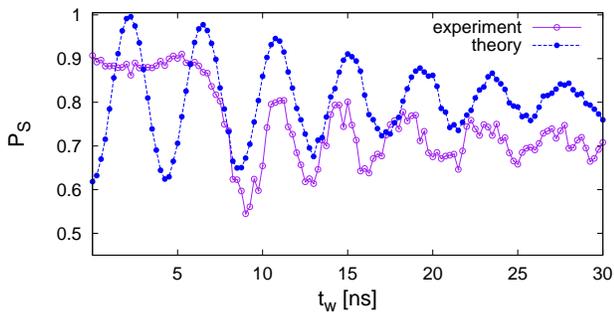}
\caption{(color online) Comparison between experimental (purple, open circles) and
theoretical values (blue, filled circles) of $P_{\mm{S}}$ for $B=45\,\mm{mT}$.
Theoretical values are obtained by finding the detuning $\varepsilon_{\mm{f}} =
3.78\,\mm{meV}$ for which the theoretical and experimental oscillation periods match.  A
suitable $\gamma = 0.12\,\mm{meV/ns}$ is then chosen such that $\varepsilon_{\mm{i}} >
\varepsilon_{\mm{c}}$. The experiments used passive filtering of square pulses to reduce
$P_{\mm{LZS}}$ and thereby increase the visibility of the oscillations. As a result,
oscillations in the experimental data are delayed to longer times due to finite rise time
effects. The theory points are obtained with a perfect square pulse.}
\label{exp_theo_fit}
\end{figure}

The model predicts coherent oscillations in $P_{\mm{S}}$  for $\varepsilon_{\mm{f}} >
\varepsilon_{\mm{c}}$, i.e. in the case where the qubit has not passed the avoided
crossing. It can be explained within finite-time LZS theory, but not with the
conventional asymptotic Landau-Zener formula. In other words, even if $t_{\mm{f}} < 0$,
we have $\left|U^{\mm{LZS}}_{ij} (t_{\mm{f}},\,t_{\mm{i}})\right|^2 > 0$, which
illustrates the non-adiabatic character of the problem. For the pulse conditions used in
Ref.~\onlinecite{petta_science2010}, oscillations are not observed for
$\varepsilon_{\mm{f}} > \varepsilon_{\mm{c}}$, most likely due to charge dephasing. The
coherence time of an admixture of $(1,\,1)$ and $(2,\,0)$ charge state has been measured
to be $\sim 1\,\mm{ns}$ for GaAs QDs\cite{hayashi_prl2003} and sets the time scale at
which the system must be driven to observe oscillations for $\varepsilon_{\mm{f}} >
\varepsilon_{\mm{c}}$. A finite-time effect in agreement with the experimental data is
the dependence of the oscillation amplitude $P_{\mm{S}}$ on the pulse length. Finally, we
show in Fig.~\ref{exp_theo_fit} a comparison between experiment and theory for
$B=45\,\mm{mT}$. The experimental data were obtained from the setup used in Ref.
\onlinecite{petta_science2010}. The experiments used passive filtering of square pulses
to reduce $P_{\mm{LZS}}$ and thereby increase the visibility of the oscillations.  As a
result, oscillations in the experimental data are delayed to longer times due to finite
rise time effects.

\section{Arbitrary single qubit rotations}

\begin{figure}
\includegraphics[width=0.48\textwidth]{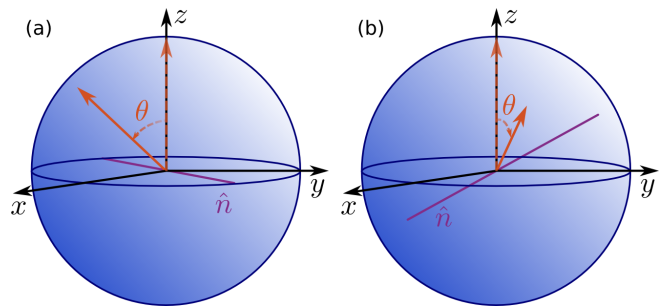}
\caption{(color online) Bloch sphere representation of LZS transitions as rotations. (a)
When the propagation times are equal, $-t_{\mm{i}}=t_{\mm{f}}$, 
the rotation axis lays in the $xy$-plane. In addition with the rotations generated around the
$z$-axis by letting the qubit evolve in the external magnetic field, any rotation can be
achieved by the Euler angle method. (b) If the propagation times are different the rotations
axis may not lay (see Appedix \ref{sec:lzsrot}) in the $xy$ plane. In this case, a pure
LZS interferometry technique can be used to generate any rotation.}
\label{single_gate}
\end{figure}

The passage through the avoided crossing can be interpreted as a rotation (see Appendix
\ref{sec:lzsrot}), $U_{\mm{LZS}} (\eta,\,t_{\mm{f}},\,t_{\mm{i}}) =
\mm{e}^{-\mm{i}\mathbf{\sigma}\cdot\mathbf{\hat{n}} \theta /2}$ by an angle $\theta$
around the axis $\mathbf{\hat{n}}$, see Fig.~ \ref{single_gate}. Here $\theta = \theta
(\eta,\,t_{\mm{f}},\,t_{\mm{i}})$ and $\mathbf{\hat{n}} =
\mathbf{\hat{n}}(\eta,\,t_{\mm{f}},\,t_{\mm{i}})$ is a unit vector where $\eta =
\left|\bra{\mm{T}_+} H_{\mm{HF}} \ket{\mm{S}}\right|/ \sqrt{\alpha \hbar}$ is the
dimensionless coupling strength. Since $\hat{n}$ and $\theta$ are functions of the same
experimental parameters $t_{\mm{i}} (\to \varepsilon_{\mm{i}}),\,t_{\mm{f}} (\to
\varepsilon_{\mm{f}}),\,\mm{and}\,\alpha (\to \gamma)$, it is not straight forward to
find them simultaneously in order to build a given single qubit rotation. For instance,
fixing two parameters and tuning the third one will simultaneously change $\hat{n}$ and
$\theta$ limiting the achievable rotations angles. Nevertheless, the situation is not
hopeless and several composite methods can be engineered to achieve any rotation. We
present here three methods, each of them having their own advantages.  

Since rotations by an angle $\varphi$ around the $z$-axis are available by letting the
qubit evolve in an external $B$ field, we would like $\mathbf{\hat{n}}$ to be in the
$xy$-plane in order to build any rotation by the Euler angle method. Below we show that
this is possible if for example (see Appendix \ref{sec:lzsrot}) the propagation times are
equal, $-t_{\mm{i}} = t_{\mm{f}} = t_{\mm{LZS}}$. However, a $\pi$-rotation from
$\ket{\mm{S}}$ to $\ket{\mm{T}_+}$ would take an exponentially long time with a single
LZS transition, since it corresponds to a fully adiabatic transition. 
%
%
However, this problem can be circumvented by sequentially applying several LZS
transitions. A $\pi$-rotation from $\ket{\mm{S}}$ to $\ket{\mm{T}_+}$ can be achieved in
$\sim 0.1 - 1.0\,\mm{ns}$ for two consecutive and identical LZS transitions.

The single qubit gates can also be implemented with a pure LZS interferometry technique,
similar to the one used to control superconducting qubits. This method requires
sequential driving of the qubit through the avoided crossing. The different passages
result in a series of LZS transitions each of them corresponding to a rotation of the
qubit. By tuning $\gamma$ and choosing $-t_{\mm{i}}\neq t_{\mm{f}}$ with different ratios
for $|t_{\mm{i}}|/|t_{\mm{f}}|$, any qubit rotations can be achieved within a
nanosecond.

Since finite-time effects are present in the system, we can think about a control method
where the qubit is operated on the $(1,\,1)$ charge configuration side
(``sweet region''). This requires the preparation of a $\mm{T}_+(1,\,1)$ state
\cite{foletti_nphys2009} and pulses with rise times shorter than $1\,\mm{ns}$ which
do not drive the system through the avoided crossing unless a measurement is required.
The qubit manipulation is achieved through finite-time LZS interferometry, where tuning
$\gamma$ and the propagation time $t_{\mm{i},\,\mm{f}}$ allows to achieve any desirable angle.
For all those methods, an arbitrary single qubit rotation can be expressed as a series of
a forward sweep~-wait~-backward sweep operators, 
\begin{equation}
\mathcal{D}(\theta,\,\varphi)=\prod_{i=1}^l
U_{\mm{b}}(\theta_{\mm{b}}^{(i)}) U_z (\varphi^{(i)}) U_{\mm{f}}(\theta_{\mm{f}}^{(i)}),
\label{rotmeth}
\end{equation}
which reduces to  Eq.~\eqref{PS_eff} for $l=1$. The proposed methods require a maximum of
$l = 3$. It is important to notice that the rotation axis and the final measurable angles
will not be $\mathbf{\hat{n}}$, $\theta = \sum_i \theta_i $, and $\varphi= \sum_i
\varphi_i$, but rather $\langle \mathbf{\hat{n}} \rangle_{\mathbf{\chi}}$, $\langle
\theta \rangle_{\mathbf{\chi}}$, and $\langle \varphi\rangle_{\mathbf{\chi}}$, where the
brackets denote the averaging over the nuclear spin bath.  A similar scheme with $l=1$
has been proposed for the $\mm{S}$ - $\mm{T}_0$ qubit \cite{hanson_prl2007}.

\section{Two qubit gate}

To complete the set of quantum gates, a two-qubit operation such as $\mm{CNOT}$
is required. We consider the Hamiltonian 
\begin{equation}
H_{\mm{eff}}^{(1)} + H_{\mm{eff}}^{(2)} + H_{\mm{int}}, 
\label{h_dgate}
\end{equation}
where $H_{\mm{eff}}^{(i)}$ is the single qubit Hamiltonian~\eqref{h_eff}
and $H_{\mm{int}} = \tilde{u}\sum_{j=2,3} n_{j\uparrow}n_{j\downarrow}$ describes the
capacitive coupling \cite{hanson_prl2007,taylor_nphys2005,stepanenko_prb2007} between two
adjacent QDs belonging to different $\mm{S}-\mm{T}_+$ qubits. 
\begin{figure}
\includegraphics[width=0.48\textwidth]{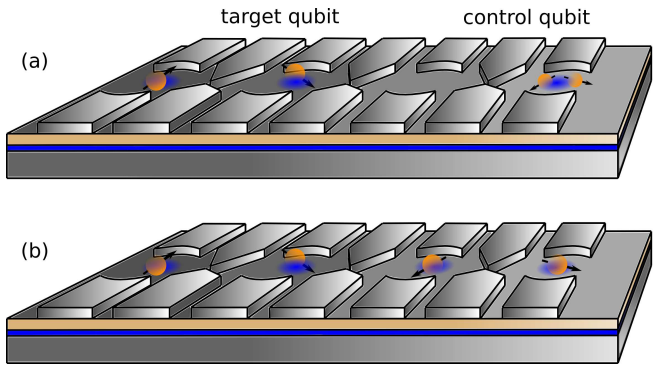}
\caption{(color online) A conditional gate can be implemented by capacitively coupling
electrons trapped in quantum dots belonging to different qubits. The crossing position
$\varepsilon_{\mm{c}}$ has different values whenever the charge state of the control
qubit is $(0\,2)$ (a) or $(1,\,1)$ (b). The later case results in $\varepsilon_{\mm{f}}
\ll \varepsilon_{\mm{c}}$ which suppresses any LZS transition.}
\label{double_gate}
\end{figure}
Tunneling between the dots
of the different qubits can be suppressed by an appropriate gate voltage. If the control
qubit, qubit-1, is in a $\mm{S}(2,\,0)$ state, $H_{\mm{int}} =0$ and the dynamics of the
target qubit, qubit-2, is reduced to the case of a single qubit, see
Fig.~\ref{double_gate}.  When the control qubit
is in a $(1,\,1)$ charge configuration, the target qubit is influenced by the interdot Coulomb
interaction. In this case, the dynamics of the target qubit can be described by
Eq.~\eqref{h_eff} by replacing $u\to u+\tilde{u}$. In particular, this affects the
position of the avoided crossing. For a system with two DQD separated by a distance $d
\simeq 2R - 10R$, where $R$ is the approximate radius of one QD, the intradot Coulomb
interaction $u \sim e^2 / R$ is comparable to the interdot Coulomb interaction $\tilde{u}
\sim e^2 / d$ resulting in $|\varepsilon_{\mm{c}} (u) - \varepsilon_{\mm{c}}
(u+\tilde{u})|>1\,\mm{meV}$. From the previous discussion, we know that a $\pi$-rotation
is possible within $\sim 1\,\mm{ns}$ if $\varepsilon_{\mm{c}} = \varepsilon_{\mm{c}}
(u)$. In the case where the avoided crossing is at $\varepsilon_{\mm{c}} (u+\tilde{u})$
the same LZS sequence will leave the target qubit unchanged, even within a finite-time
theory since the separation between the two avoided crossings is $> 1\,\mm{meV}$.
Therefore, we estimate the $\mm{CNOT}$ gate time to be $\sim 1 - 3\,\mm{ns}$. 

Let us consider the case where the control qubit is in the $\mm{S}(2,\,0)$ state, which
is the logical $\ket{1}$ of the qubit,  such that the target and control qubit are not
capacitively coupled, $u=4\,\mm{meV}$. For $\gamma = 0.3\,\frac{\mm{meV}}{\mm{ns}}$,
$B=100\,\mm{mT}$, $\varepsilon_{\mm{i}} = 3.98597\,\mm{meV}$, and $\varepsilon_{\mm{f}}
=3.97991\,\mm{meV}$ we find for $l=6$ and $\varphi^{(i)} = 0$ in Eq.~\eqref{rotmeth} 
\begin{equation}
\mathcal{D}^{(1)}(\theta,\,\varphi) \simeq
\begin{pmatrix}
-0.314 &&  0.945 - 0.096\ui \\
-0.945 - 0.096\ui &&  -0.314
\end{pmatrix}.
\label{example_cqubit=1}
\end{equation}
This example shows the almost perfect realization of a conditional 
 $\ui \sigma_y$ operation which corresponds to
a CNOT gate up to single-qubit gates. 

To show that this method produces a CNOT gate, we consider the case where
the control qubit is in the $\mm{T}_+ (1,\,1)$ state, which is the logical $\ket{0}$ of
the qubit.
We estimate a lower bound for the strength of the capacitive coupling 
between the qubits to be $u + \tilde{u} = 5\,\mm{meV}$ (see above).
In this case, the target qubit evolution takes the form 
\begin{equation}
\mathcal{D}^{(0)}(\theta,\,\varphi) \simeq 
\begin{pmatrix}
0.971 + 0.002\ui && -0.206 - 0.124\ui \\
0.206 - 0.124\ui && 0.971 - 0.002\ui
\end{pmatrix},
\label{example_cqubit=0}
\end{equation}
which is close to $\mathbbm{1}$ and demonstrates the possibility of generating a CNOT
gate with the proposed method.

Notice that our choice for $\varepsilon_{\mm{i,\,f}}$ corresponds
to a propagation time $t_{\mm{LZS}} = 0.01\,\mm{ns}$ such that the total gate time to
achieve controlled-$\ui \sigma_y$ is $0.24\,\mm{ns}$. 
The fidelity is $\mathcal{F} =
\abs{\mm{Tr}\,\mathcal{D}^{(0)}+ \mm{Tr}\left[\mathcal{D}^{(1)}(\ui\sigma_y)^{\dag}\right]}^2/16 \simeq
 0.918$.
A more accurate CNOT gate can be engineered by fine tuning the parameters entering the
LZS propagator. 

\section{Conclusions}

We have demonstrated that coherent control of the $\mm{S}$-$\mm{T}_+$ qubit can be
achieved using LZS interferometry. Hyperfine interactions lead to an avoided crossing
between $\mm{S}$ and $\mm{T}_+$ states, which allows for efficient quantum control.
Moreover, we predict that in the limit of fast rise-time pulses coherent oscillations in
$P_{\mm{S}}$ should be observed even without going through the avoided crossing. This
phenomenon is a finite time effect which we have theoretically described using the 
general finite-time LZS theory and it can be used to operate the qubit in the $(1,\,1)$
charge configuration side (``sweet region'').

Our scheme can be extended to DQD in materials with few nuclear spins (graphene,
CNT, Si). In such cases, the avoided crossing between the qubit states can be achieved by
engineering a DQD in the presence of micro-magnets which provide the in-plane gradient
magnetic field for the realization of the LZS based gates \cite{micromagnet}. The qubit
will moreover benefit from the lack of the inhomogeneous broadening due to the Overhauser
fields and exhibit an extended $T_2^{*}$. In GaAs DQDs the method proposed in
Ref.~\onlinecite{foletti_nphys2009} could be used to extend $T_2^{*}$ without cancelling the
gradient field. Other schemes to polarize the nuclear spins \cite{petta_prl2008} or
reduce their fluctuations \cite{klauser_coish_prb2006,stepanenko_prl2006} also exist.


\section{Acknowledgments}

We acknowledge funding
from the DFG within SPP 1285, FOR 912 and SFB 767.  Research at Princeton
was supported by the Sloan and Packard Foundations, DARPA award N66001-09-1-2020, and the NSF
through DMR-0819860 and DMR-0846341.

\appendix 

\section{The Landau-Zener-Stückelberg finite-time propagator}
\label{sec:lzsprop}

In this appendix we follow the work of Vitanov and Garraway \cite{vitanov_time_lz} and
consider, without loss of generality, a two-level system whose eigenenergies $E_1$ and
$E_2$ are time dependent, $E_1 = E_1 (t),\,E_2 = E_2 (t)$ and their difference is a
linear function of time $2 \Delta (t) = E_2 (t) - E_1(t) =\alpha t$. Furthermore, we
assume the levels to be coupled with strength $\lambda$. The matrix representation of the
system's Hamiltonian is given by
\begin{equation}
H_{\mm{LZS}} (t)= 
\begin{pmatrix}
-\Delta (t) && \lambda \\
\lambda && \Delta (t)
\end{pmatrix}
.
\label{hlzs}
\end{equation}

The time evolution of such a system is described by the time-dependent Schrödinger
equation 

\begin{equation}
\ui \hbar \drv{}{t}\ket{\psi (t)} = H_{\mm{LZS}} (t)\ket{\psi (t)}
\label{sch_eq}
\end{equation}
with $\ket{\psi (t)} = c_1 (t) \ket{1} + c_2 (t) \ket{2}$. After substitution of
\eqref{hlzs} into Eq.~\eqref{sch_eq}, a coupled system of first order ordinary
differential equations is obtained 

\begin{empheq}{align}
\ui \hbar \dot{c}_1 (t) & = -\Delta (t) c_1 (t) + \lambda c_2 (t),
\label{lzs_diffeq_syst:1}\\
\ui \hbar \dot{c}_2 (t) & = \lambda c_1 (t) + \Delta (t) c_2 (t).
\label{lzs_diffeq_syst:2}
\end{empheq}

By deriving Eq.~\eqref{lzs_diffeq_syst:1} with respect to time and substituting
Eqs.~\eqref{lzs_diffeq_syst:1}~and~\eqref{lzs_diffeq_syst:2} into the newly obtained
ordinary second order differential equation, we obtain 

\begin{equation}
\ddot{c}_1 (t) = \left( \frac{\ui}{\hbar} \alpha - \frac{\alpha^2 t^2}{\hbar^2} -
\frac{\lambda^2}{\hbar^2}\right)c_1 (t).
\label{weber_diff_eq}
\end{equation}

It is convenient to introduce dimensionless parameters before solving
Eq.~\eqref{weber_diff_eq}, here we introduce the dimensionless time $\tau =
\sqrt{\frac{\alpha}{\hbar}}t$ which we substitute in Eq.~\eqref{weber_diff_eq} to
obtain 

\begin{equation}
\drv{^2}{\tau^2} c_1 (\tau) + \left( -\ui + \eta^2 + \tau^2\right)c_1 (\tau) = 0,
\label{dimless_weber_diff_eq}
\end{equation}
where $\eta = \frac{\lambda}{\sqrt{\alpha \hbar}}$ is the dimensionless coupling
strength. 

The solution of Eq.~\eqref{dimless_weber_diff_eq} is 
\begin{equation}
c_1 (t) = \kappa_1 D_{\frac{\ui \eta^2}{2}}\left(\sqrt{2} \ue^{-\frac{\ui \pi
}{4}}\tau\right) + \kappa_2 D_{\frac{\ui \eta^2}{2}}\left(\sqrt{2} \ue^{\frac{3 \ui \pi
}{4}}\tau\right),
\label{c1_weber_dimless}
\end{equation}
where $D_\nu (z)$ are parabolic cylinder functions, which solve the Weber
equation\cite{abramowitz}
\begin{equation} 
\drv{^2}{z^2}D_\nu (z) + \left(\nu + \frac{1}{2} - \frac{1}{4}z^2\right)D_\nu (z) = 0,
\label{weber_eq} 
\end{equation}
and can be obtained from Eq.~\eqref{dimless_weber_diff_eq} by writing the expression in
brackets as $-2\ui (\ui \eta^2 / 2 + 1/2 + \ui \tau^2 /2)$ and substituting $\tau \to
2^{-1/2} \exp(\ui \pi /4) z$. 

$c_2 (t)$ is obtained by inserting Eq.~\eqref{c1_weber_dimless} into
Eq.~\eqref{lzs_diffeq_syst:1} and using the property 
\begin{equation}
\drv{}{z}\left(\ue^{\frac{z^2}{4}}D_\nu (z)\right) = \nu \ue^{\frac{z^2}{4}} D_{\nu -1}
(z).
\label{derivative_D}
\end{equation}
One finds
\begin{equation}
\begin{split}
c_2 (t) = \frac{\eta}{\sqrt{2}} \ue^{-\frac{\ui \pi} {4}}\left[-\kappa_1 D_{\frac{\ui
\eta^2}{2} - 1}\left(\sqrt{2} \ue^{-\frac{\ui\pi}{4}}\tau\right)\right.\\ 
\left.+ \kappa_2 D_{\frac{\ui \eta^2}{2} - 1}\left(\sqrt{2}
\ue^{\frac{3\ui\pi}{4}}\tau\right)\right].
\label{c2_weber_dimless}
\end{split}
\end{equation}

To find the constants $\kappa_1$ and $\kappa_2$, we consider initial conditions given by $c_1
(\tau_{\mm{i}})$ and $c_2 (\tau_{\mm{i}})$ and the Wronskian relation 
\begin{equation}
\begin{split}
\mm{W}\left\{D_\nu (z),\,D_\nu (-z)\right\} &\coloneqq D_\nu (z) \drv{}{z}D_\nu (-z) -
D_\nu (-z) \drv{}{z}D_\nu (z) \\
&= \frac{\sqrt{2 \pi}}{\Gamma (-\nu)}.
\end{split}
\label{wronskian_para_cyl}
\end{equation}
We solve the system of equation given by
Eqs.~\eqref{c1_weber_dimless}~and~\eqref{c2_weber_dimless} for $\kappa_1$ and $\kappa_2$
using the Wronskian property~\eqref{wronskian_para_cyl},
we find 
\begin{empheq}{align}
\kappa_1 &= \frac{\Gamma\left(1 - \frac{\ui \eta^2}{2}\right)}{\sqrt{2\pi}} \left[
D_{\frac{\ui \eta^2}{2} - 1}\left(\sqrt{2} \ue^{\frac{3\ui\pi}{4}}\tau_{\ui}\right)c_1
(\tau_{\ui})\right.\notag \\
&\phantom{=} \left.- \frac{\sqrt{2}}{\omega} \ue^{\frac{\ui \pi}{4}}D_{\frac{\ui
\eta^2}{2}}\left(\sqrt{2} \ue^{\frac{3\ui\pi}{4}}\tau_{\ui}\right)c_2
(\tau_{\ui})\right], \label{const_diff_eq:1}\\
\kappa_2 &= \frac{\Gamma\left(1 - \frac{\ui \eta^2}{2}\right)}{\sqrt{2\pi}} \left[
D_{\frac{\ui \eta^2}{2} - 1}\left(\sqrt{2} \ue^{-\frac{\ui\pi}{4}}\tau_{\ui}\right)c_1
(\tau_{\ui})\right. \notag \\
&\phantom{=} \left.+ \frac{\sqrt{2}}{\omega} \ue^{\frac{\ui \pi}{4}}D_{\frac{\ui
\eta^2}{2}}\left(\sqrt{2} \ue^{-\frac{\ui\pi}{4}}\tau_{\ui}\right)c_2
(\tau_{\ui})\right]. \label{const_diff_eq:2}
\end{empheq}

Substituting Eqs.~\eqref{const_diff_eq:1}~and~\eqref{const_diff_eq:2} into
Eqs.~\eqref{c1_weber_dimless}~and~\eqref{c2_weber_dimless} and having in mind that we are
looking for the evolution operator $U(t_{\mm{f}},\,t_{\mm{i}})$ giving the final state
knowing the initial one
\begin{equation}
\ket{\psi (t_{\mm{f}})} = U(t_{\mm{f}},\,t_{\mm{i}})\ket{\psi (t_{\mm{i}})}
\label{evo_op}
\end{equation}
we finally find the LZS propagator 
\begin{equation}
U_{\mm{LZS}}(t_{\mm{f}},\,t_{\mm{i}}) =
\begin{pmatrix}
u_{11} (t_{\mm{f}},\,t_{\mm{i}})  & u_{12}(t_{\mm{f}},\,t_{\mm{i}})\\
u_{21}(t_{\mm{f}},\,t_{\mm{i}}) & u_{22}(t_{\mm{f}},\,t_{\mm{i}}) 
\end{pmatrix}
\label{lzs_matrix}
\end{equation}
with 
\begin{equation}
\begin{split}
&u_{11} (t_{\mm{f}},\,t_{\mm{i}}) = u_{22}^{\ast}(t_{\mm{f}},\,t_{\mm{i}})=\\
& \frac{\Gamma\left(1 - \frac{\ui\eta^2}{2}\right)}{\sqrt{2\pi}} 
\left[D_{\frac{\ui \eta^2}{2}}\left(\sqrt{2} \ue^{-\frac{\ui\pi}{4}}\tau_{\mm{f}}\right)
D_{\frac{\ui\eta^2}{2}-1}\left(\sqrt{2}\ue^{\frac{3\ui\pi}{4}}\tau_{\mm{i}}\right)
\right.\\
&+ \left. D_{\frac{\ui \eta^2}{2}}\left(\sqrt{2}
\ue^{\frac{3\ui\pi}{4}}\tau_{\mm{f}}\right)
D_{\frac{\ui\eta^2}{2}-1}\left(\sqrt{2}\ue^{-\frac{\ui\pi}{4}}\tau_{\mm{i}}\right)
\right]
\end{split}
\label{u11}
\end{equation}
and
\begin{equation}
\begin{split}
&u_{12} (t_{\mm{f}},\,t_{\mm{i}}) = -u_{21}^{\ast}(t_{\mm{f}},\,t_{\mm{i}})=\\
& \frac{\Gamma\left(1 - \frac{\ui\eta^2}{2}\right)}{\sqrt{\pi}\eta}
\ue^{\frac{\ui\pi}{4}}
\left[-D_{\frac{\ui \eta^2}{2}}\left(\sqrt{2} \ue^{-\frac{\ui\pi}{4}}\tau_{\mm{f}}\right)
D_{\frac{\ui\eta^2}{2}}\left(\sqrt{2}\ue^{\frac{3\ui\pi}{4}}\tau_{\mm{i}}\right)
\right.\\
&+ \left. D_{\frac{\ui \eta^2}{2}}\left(\sqrt{2}
\ue^{\frac{3\ui\pi}{4}}\tau_{\mm{f}}\right)
D_{\frac{\ui\eta^2}{2}}\left(\sqrt{2}\ue^{-\frac{\ui\pi}{4}}\tau_{\mm{i}}\right)
\right].
\end{split}
\label{u12}
\end{equation}
In the original LZS problem $t=0$ is defined at the energy levels crossing. A situation
where $t_{\mm{i}} < 0$ and $t_{\mm{f}} > 0$ corresponds to drive the system through the
avoided crossing. The case $t_{\mm{i}} < 0$ and $t_{\mm{f}} < 0$ corresponds to stop the
system before it goes through the avoided crossing. Finally, $t_{\mm{i}} > 0$ and
$t_{\mm{f}} > 0$ corresponds to a system which is initially prepared after the avoided
crossing.

\section{Asymptotic expansion of the parabolic cylinder functions}

The expression of the LZS propagator can be expressed with simpler functions when the
argument $\tau \gg 1$ and the parameter $\eta \gg 1$, in this case the parabolic cylinder
functions can be expanded asymptotically\cite{olver_1959}. The necessary
asymptotic forms to expand Eqs.~\eqref{u11}~and~\eqref{u12} are 
\begin{equation}
D_{\frac{\pm\ui\eta^2}{2}}\left(\sqrt{2}\ue^{\mp\frac{\ui\pi}{4}}\tau\right) \simeq
\cos\theta\ue^{\frac{\pi \eta^2}{8} \pm \ui \xi}\,,
\label{exp:1}
\end{equation}


\begin{equation}
D_{\frac{\pm\ui\eta^2}{2}-1}\left(\sqrt{2}\ue^{\frac{\mp\ui\pi}{4}}\tau\right) \simeq
\frac{\sqrt{2}}{\omega}\sin\theta\ue^{\frac{\pi \eta^2}{8} \pm \ui (\xi + \frac{\pi}{4})}\,,
\label{exp:3}
\end{equation}


\begin{equation}
\begin{split}
D_{\frac{\ui\eta^2}{2}}\left(\sqrt{2}\ue^{\frac{3\ui\pi}{4}}\tau\right) &\simeq
\cos\theta\ue^{-\frac{3 \pi \eta^2}{8} + \ui \xi} \\
&+ \frac{\eta \sqrt{\pi}}{\Gamma\left(1
- \frac{\ui \eta^2}{2}\right)}\sin\theta\ue^{-\frac{\pi \eta^2}{8}-\ui(\xi +
\frac{\pi}{4})}\,,
\end{split}
\label{exp:5}
\end{equation}

\begin{equation}
\begin{split}
D_{\frac{\ui\eta^2}{2}-1}\left(\sqrt{2}\ue^{\frac{3\ui\pi}{4}}\tau\right) &\simeq
\frac{\sqrt{2}}{\eta}\sin\theta\ue^{-\frac{3 \pi \eta^2}{8} + \ui (\xi-\frac{3\pi}{4})} \\
&+ \frac{\sqrt{2 \pi}}{\Gamma\left(1
- \frac{\ui \eta^2}{2}\right)}\cos\theta\ue^{-\frac{\pi \eta^2}{8}-\ui\xi}\,,
\end{split}
\label{exp:6}
\end{equation}
where we have defined 

\begin{equation}
\xi = -\frac{\eta^2}{4}+\frac{\eta^2}{2}\ln\left(\frac{1}{\sqrt{2}}(\tau + \sqrt{\tau^2 +
\eta^2})\right)+\frac{\tau}{2}\sqrt{\tau^2 + \eta^2}\,,
\label{xi}
\end{equation}
and 

\begin{equation}
\begin{aligned}
\cos\theta &= \sqrt{\frac{1}{2}\left(1 + \frac{\tau}{\sqrt{\tau^2 + \eta^2}}\right)},\\
\sin\theta &= \sqrt{\frac{1}{2}\left(1 - \frac{\tau}{\sqrt{\tau^2 + \eta^2}}\right)}.
\end{aligned}
\label{sincos}
\end{equation}

Using the above expressions and writing $\tau_{\mm{i}} = \ue^{\ui \pi}
\abs{\tau_{\mm{i}}}$ to fulfill the condition for the expansion we find, 

\begin{equation}
\begin{aligned}	
&u_{11}(t_{\mm{f}},\,t_{\mm{i}}) = u_{22}^{\ast}(t_{\mm{f}},\,t_{\mm{i}}) \simeq \\
&\sqrt{1 - \ue^{-\pi \eta^2}}\left( \sin\theta_{\mm{f}}\cos\theta_{\mm{i}}\ue^{-\ui
\left(\xi_{\mm{f}} + \xi_{\mm{i}} + \mm{arg}\Gamma\left(1- \frac{\ui \eta^2}{2}\right) +
\frac{\pi}{4}\right)}\right.\\
&\left.\phantom{\sqrt{1 - \ue^{-\pi \eta^2}}(} 
+ \sin\theta_{\mm{i}}\cos\theta_{\mm{f}}\ue^{\ui
\left(\xi_{\mm{f}} + \xi_{\mm{i}} + \mm{arg}\Gamma\left(1- \frac{\ui \eta^2}{2}\right) +
\frac{\pi}{4}\right)}\right)\\
& + \ue^{-\frac{\pi\eta^2}{2}}
\left(\cos\theta_{\mm{f}}\cos\theta_{\mm{i}}\ue^{\ui(\xi_{\mm{f}} - \xi_{\mm{i}})} -
\sin\theta_{\mm{f}}\sin\theta_{\mm{i}}\ue^{-\ui(\xi_{\mm{f}} - \xi_{\mm{i}})}\right),
\end{aligned}
\label{u11_asympt}
\end{equation}
and 

\begin{equation}
\begin{aligned}
&u_{12}(t_{\mm{f}},\,t_{\mm{i}}) = -u_{21}^{\ast}(t_{\mm{f}},\,t_{\mm{i}}) \simeq \\
&\sqrt{1 - \ue^{-\pi \eta^2}}\left( \sin\theta_{\mm{f}}\sin\theta_{\mm{i}}\ue^{-\ui
\left(\xi_{\mm{f}} + \xi_{\mm{i}} + \mm{arg}\Gamma\left(1- \frac{\ui \eta^2}{2}\right)
+ \frac{\pi}{4}\right)}\right.\\
&\left.\phantom{\sqrt{1 - \ue^{-\pi \eta^2}}(} 
- \cos\theta_{\mm{f}}\cos\theta_{\mm{i}}\ue^{\ui
\left(\xi_{\mm{f}} + \xi_{\mm{i}} + \mm{arg}\Gamma\left(1- \frac{\ui \eta^2}{2}\right) +
\frac{\pi}{4}\right)}\right)\\
& + \ue^{-\frac{\pi\eta^2}{2}}
\left(\cos\theta_{\mm{i}}\sin\theta_{\mm{f}}\ue^{-\ui(\xi_{\mm{f}} - \xi_{\mm{i}})} +
\cos\theta_{\mm{f}}\sin\theta_{\mm{i}}\ue^{\ui(\xi_{\mm{f}} - \xi_{\mm{i}})}\right).
\end{aligned}
\label{u12_asympt}
\end{equation}
where $\cos\theta_{\mm{i,\,f}},\,\sin\theta_{\mm{i,\,f}}$, and $\xi_{\mm{i,\,f}}$ are
respectively given by Eq.~\eqref{sincos} and Eq.~\eqref{xi} for $\tau =
\ue^{\ui\pi}\abs{\tau_{\mm{i}}},\,\tau_{\mm{f}}$.

We noticed that the asymptotic expansions \eqref{exp:1}, \eqref{exp:3}, \eqref{exp:5},
and \eqref{exp:6} are valid for the weaker condition $\tau + \eta \gg 1$, as already
reported in Ref.~\onlinecite{vitanov_time_lz} for the expansion of $\abs{u_{11}}^2$ and
$\abs{u_{12}}^2$.

\section{The LZS propagator as a rotation}
\label{sec:lzsrot}

In quantum mechanics the rotation operator $\mathcal{D}(\hat{n},\,\vartheta)$ by an angle
$\vartheta$ around an axis $\hat{n}$ of a two-level system has the representation 
\begin{equation}
\begin{aligned}
\mathcal{D}(\hat{n},\,\vartheta) &=
\ue^{\ui\hat{n}\cdot\mathbf{\sigma}\frac{\vartheta}{2}}\\
&=\begin{pmatrix}
\cos\frac{\vartheta}{2} - \ui n_z \sin\frac{\vartheta}{2} & (-\ui n_x - n_y)
\sin\frac{\vartheta}{2}\\
(-\ui n_x + n_y)\sin\frac{\vartheta}{2} & \cos\frac{\vartheta}{2} + \ui n_z
\sin\frac{\vartheta}{2}
\end{pmatrix}.
\end{aligned}
\label{rotation}
\end{equation}

Identifying Eqs.~\eqref{rotation}~and~\eqref{lzs_matrix} with $u_{ij}$ given by
Eqs.~\eqref{u11_asympt}~and~\eqref{u12_asympt} we can express the rotation angle
$\vartheta$ and the rotation axis $\hat{n}$ as functions of the LZS propagator parameters
$\tau_{\mm{i}},\,\tau_{\mm{f}},\,\eta$. We have 

\begin{equation}
\begin{aligned}
\cos\frac{\vartheta}{2} &= \sqrt{1 - \ue^{-\pi \eta^2}}
\cos\zeta\left[\sin\theta_{\mm{f}}\cos\theta_{\mm{i}} +
\cos\theta_{\mm{f}}\sin\theta_{\mm{i}}\right]\\
&\phantom{=}+ \ue^{-\frac{\pi\eta^2}{2}} \cos\varsigma\left[\cos\theta_{\mm{f}}\cos\theta_{\mm{i}} -
\sin\theta_{\mm{f}}\sin\theta_{\mm{i}}\right],
\end{aligned}
\label{costheta}
\end{equation}
where 

\begin{equation}
\zeta = \xi_{\mm{f}} + \xi_{\mm{i}} + \mm{arg}\Gamma\left(1 - \frac{\ui
\eta^2}{2}\right)+\frac{\pi}{4},
\label{zeta}
\end{equation}
and 

\begin{equation}
\varrho = \xi_{\mm{f}}-\xi_{\mm{i}}.
\label{varsigma}
\end{equation}
%
%
%

\begin{figure}
\includegraphics[width=0.48\textwidth]{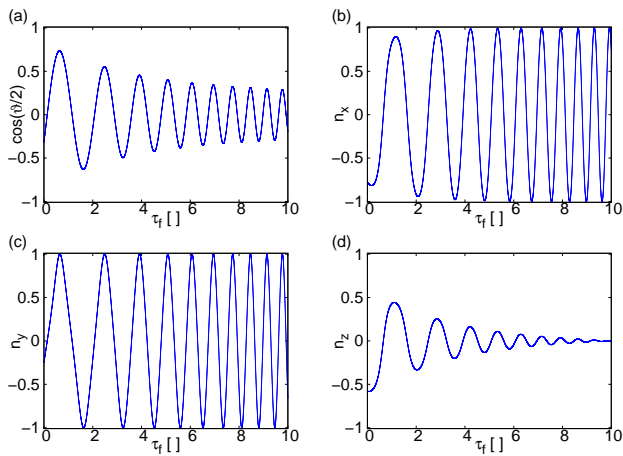}
\caption{(color online) (a) Cosine of the rotation angle and components of the rotation
axis (b), (c), and (d)
as a function of $\tau_{\mm{f}}$ for the dimensionless parameters $\tau_{\mm{i}}=10$ and
$\eta=3$.}
\label{nandcos}
\end{figure}

The components of the rotation axis are given by (see Fig.~\ref{nandcos})

\begin{equation}
\begin{aligned}
n_x &= \frac{1}{\sin\frac{\vartheta}{2}} \left(\sqrt{1
-\ue^{-\pi\eta^2}}\sin\zeta\left[\sin\theta_{\mm{f}}\sin\theta_{\mm{i}}+\cos\theta_{\mm{f}}\cos\theta_{\mm{i}}\right]\right.\\
&\phantom{=\frac{1}{\sin\frac{\vartheta}{2}} (} \left.+ \ue^{-\frac{\pi
\eta^2}{2}}\sin\varrho\left[\cos\theta_{\mm{i}}\sin\theta_{\mm{f}} -
\cos\theta_{\mm{f}}\sin\theta_{\mm{i}}\right]\right),
\end{aligned}
\label{nx}
\end{equation}

\begin{equation}
\begin{aligned}
n_y &= \frac{1}{\sin\frac{\vartheta}{2}} \left(\sqrt{1
-\ue^{-\pi\eta^2}}\cos\zeta\left[-\sin\theta_{\mm{f}}\sin\theta_{\mm{i}} +
\cos\theta_{\mm{f}}\cos\theta_{\mm{i}}\right]\right.\\
&\phantom{\frac{1}{\sin\frac{\vartheta}{2}} (} \left.+ \ue^{-\frac{\pi
\eta^2}{2}}\cos\varrho\left[-\cos\theta_{\mm{i}}\sin\theta_{\mm{f}} -
\cos\theta_{\mm{f}}\sin\theta_{\mm{i}}\right]\right),
\end{aligned}
\label{ny}
\end{equation}

\begin{equation}
\begin{aligned}
n_z &=\frac{1}{\sin\frac{\vartheta}{2}} \left(\sqrt{1
-\ue^{-\pi\eta^2}}\sin\zeta\left[\sin\theta_{\mm{f}}\cos\theta_{\mm{i}} -
\cos\theta_{\mm{f}}\sin\theta_{\mm{i}}\right]\right.\\
&\phantom{\frac{1}{\sin\frac{\vartheta}{2}} (} \left.+ \ue^{-\frac{\pi
\eta^2}{2}}\sin\varrho\left[-\cos\theta_{\mm{f}}\cos\theta_{\mm{i}} +
\sin\theta_{\mm{f}}\sin\theta_{\mm{i}}\right]\right).
\end{aligned}
\label{nz}
\end{equation}

\end{document}